\documentclass{ws-ijmpa}

\begin{document}

\markboth{E. R. Bezerra de Mello \& A. A. Saharian}
{Vacuum Polarization by Flat Boundary}

%
\catchline{}{}{}{}{}
%

\title{\bf VACUUM POLARIZATION IN A COSMIC STRING SPACETIME INDUCED BY FLAT BOUNDARY
}

\author{EUG\^ENIO R. BEZERRA DE MELLO\footnote{emello@fisica.ufpb.br}}

\address{Departamento de F\'{\i}sica-CCEN, Universidade Federal da Para\'{\i}ba\\
J. Pessoa, PB, 58.059-970, Brazil}

\author{ARAM A. SAHARIAN\footnote{saharian@ysu.am}}

\address{Department of Physics, Yerevan State University\\
1 Alex Manoogian Street, 0025 Yerevan, Armenia}

\maketitle

\begin{history}
\received{Day Month Year}
\revised{Day Month Year}
\end{history}

\begin{abstract}
In this paper we analyze the vacuum expectation values of the field squared and the energy-momentum tensor associated to a massive scalar field in a higher dimensional cosmic string spacetime, obeying Dirichlet or Neumann boundary conditions on the surface orthogonal to the string.
\keywords{Vacuum polarization; cosmic string; scalar field.}
\end{abstract}

\ccode{PACS numbers: 98.80.Cq, 11.10.Gh, 11.27.+d}

\section{Introduction}	
Cosmic strings are linear topologically stable gravitational defects which appear in the framework of grand unified theories. These objects could be produced in very early Universe as consequence of a vacuum phase transition.\cite{Kibble,V-S} The spacetime produced by an idealized cosmic string is locally flat, however globally conical, with a planar angle deficit determined by the string tension. This conical structure provides nonzero vacuum expectation values (VEVs) for different physical observables. In this context, the VEVs of the energy-momentum tensor have been calculated for scalar\cite{scalar}\cdash\cite{scalar4} and fermionic\cite{ferm}\cdash\cite{ferm3} fields. Another type of vacuum polarization arises in the presence of boundaries. In this sense, the imposed boundary conditions on quantum fields alter the zero-point fluctuation spectrum and result in additional shifts in the VEVs of physical quantities, such as the energy density and stresses. This is the well-known Casimir effect.\footnote{See Ref.~\refcite{Cas} for a review about the Casimir effect.} The analysis of the Casimir effect in the idealized cosmic string spacetime have been developed for scalar,\cite{Mello} vector \cite{Brev95,Mello1} and fermionic fields,\cite{Aram1,Beze10} respectively, obeying boundary conditions on cylindrical surfaces. Continuing along this line of investigation, here we shall analyze the contribution on the vacuum polarization effects, in a higher-dimensional cosmic string spacetime, induced by a scalar field obeying Dirichlet or Neumann boundary conditions on a flat surface orthogonal to the string.

\section{Green Function}

This section is devoted to calculate the Green function associated with a massive scalar quantum field propagating in a $D$-dimensional cosmic string spacetime. Adopting the coordinate system $x^{i}=(t,r,\varphi ,z,x^{l})$, with $\varphi\in \lbrack 0,\ 2\pi ]$, and $t,\ z,\ x^{l}\in (-\infty ,\ \infty )$, the line element is given by
\begin{equation}
ds^{2}=g_{\mu\nu}dx^{\mu}dx^{\nu}=-dt^{2}+dr^{2}+\alpha ^{2}r^{2}d\varphi^{2}+dz^{2}+\sum_{l=4}^{D-1}(dx^{l})^{2}\ .  \label{cs}
\end{equation}
The cosmic string is located on the subspace defined by $r=0$, being $r\geqslant 0$ the radial polar coordinate. In (\ref{cs}) the parameter $\alpha $, smaller than unity, codifies the presence of the string.

The equation which governs the propagation of a massive scalar field in an arbitrary curved spacetime has the form:
\begin{equation}
\left( \Box -m^{2}-\xi R\right) \phi (x)=0\ ,  \label{eq1}
\end{equation}%
with $\Box $ denoting the covariant d'Alembertian and $R$ is the scalar curvature. We have introduced in (\ref{eq1}) an arbitrary curvature coupling $\xi$. Moreover, we shall assume that the field obeys Dirichlet boundary condition on the hypersurface orthogonal to the string and located at $z=0$:
\begin{equation}
\phi (x)=0\ ,\ z=0\ .  \label{bc}
\end{equation}

The corresponding Green function should obey the differential equation,
\begin{equation}
\left( \Box -m^{2}-\xi R\right) G(x,x^{\prime })=-\frac{\delta ^{D}(x-x^{\prime })}{\sqrt{-g}}\ ,  \label{Green1}
\end{equation}
besides the boundary condition (\ref{bc}).

In order to provide an explicit form for this Green function, we shall use the complete set of solutions of the eigenvalues equation,
\begin{equation}
\left( \Box -m^{2}-\xi R\right) \Phi _{\sigma }(x)=-\sigma ^{2}\Phi _{\sigma}(x)\ , \label{eigen}
\end{equation}
with $\sigma^2\geq0$. So we may write
\begin{equation}
G(x,x^{\prime })=\int_{0}^{\infty }ds\ {\mathcal{K}}(x,x^{\prime };s)\ ,
\label{heat}
\end{equation}
where the heat kernel, ${\mathcal{K}}(x,x^{\prime };s)$, can be expressed in terms of this normalized complete set of eigenfunctions as follows:
\begin{equation}
{\mathcal{K}}(x,x^{\prime };s)=\sum_{\sigma^2}\Phi _{\sigma }(x)\Phi _{\sigma}^{\ast }(x^{\prime })e^{-s\sigma ^{2}}\ .  \label{heat1}
\end{equation}
The complete set of normalized solutions of (\ref{eigen}), compatible with the boundary condition (\ref{bc}), can be specified in terms of a set of quantum numbers $(\omega ,\ q,\ n,\ k_{z},\ k_{l})$, where $n=0,\ \pm 1,\ \pm 2,\ ...\ $, $(\omega ,\ k_{l})\in \ (-\infty ,\ \infty )$ and $(q,\ k_{z})\geqslant 0$. These functions are given by:
\begin{equation}
\Phi _{\sigma }(x)=2\sqrt{\frac{q}{\alpha }}\frac{e^{i(n\varphi +\mathbf{k}\cdot \mathbf{x}-\omega t)}}{(2\pi )^{(D-1)/2}}J_{|n|/\alpha }(qr)\sin(k_{z}z)\ ,  \label{sol}
\end{equation}%
being $J_{\nu }(x)$ the Bessel function and $\mathbf{x}=(x^{4},\ldots,x^{D-1})$. The corresponding positively defined eigenvalue is given below:
\begin{equation}
\sigma ^{2}=\omega ^{2}+q^{2}+k_{z}^{2}+\mathbf{k}^{2}+m^{2}\ .
\end{equation}

The heat kernel can be given in terms of the above eigenfunctions according to (\ref{heat1}). After performing the integrals with the help of Ref.~\refcite{Grad} we obtain:
\begin{equation}
{\mathcal{K}}(x,x^{\prime };s)=\frac{2e^{-\frac{\Delta \rho ^{2}}{4s}-sm^{2}}}{\alpha (4s\pi )^{D/2}}\sinh \left( \frac{zz^{\prime }}{2s}\right)\sum_{n=-\infty }^{+\infty }e^{in\Delta \varphi }I_{|n|/\alpha }\left( \frac{rr^{\prime }}{2s}\right) \ ,  \label{heat2}
\end{equation}%
where $I_{\nu }(x)$ is the modified Bessel function and
\begin{equation}
\Delta \rho ^{2}=-\Delta t^{2}+r^{2}+r^{\prime }{}^{2}+\Delta \mathbf{x}^{2}+z^{2}+z^{\prime }{}^{2}\ ,
\end{equation}
with $\Delta \varphi =\varphi -\varphi ^{\prime }$, $\Delta t=t-t^{\prime }$, $\Delta \mathbf{x=x-x}^{\prime }$.

The next step is to integrate the heat kernel according to (\ref{heat}), to provide a closed expression for the Green function. Unfortunately, in general case, this is not possible. Only for massless fields and for specific values of the parameter $\alpha$, the corresponding Green functions can be expressed in terms of a finite sum of the associated Legendre functions and the Macdonald ones, respectively. 

The case of a scalar field with Neumann boundary condition, $\partial _{z}\phi =0$ at $z=0$, can be considered in a similar way. The corresponding eigenfunctions have the form (\ref{sol}) with the replacement $\sin (k_{z}z)\rightarrow \cos (k_{z}z)$. The expression for the heat kernel is obtained from (\ref{heat2}) with the replacement $\sinh (zz^{\prime }/(2s))\rightarrow \cosh(zz^{\prime }/(2s))$.

\subsection{Special case}
The analysis of vacuum polarization effects associated with a quantum scalar field in a cosmic string spacetime when parameter $\alpha $ is equal to the inverse of an integer number $p$, i.e., when $\alpha =1/p$, have been considered by many authors.\cite{scalar1}-\cite{scalar3} Although being a very special situation, the corresponding analysis may shed light on the qualitative behavior of these quantities for non-integer $p$. For this case, the expression of the heat kernel can be simplified with the help of the formula:\cite{Pru,Jean}
\begin{equation}
\sum_{n=-\infty }^{+\infty }e^{in\Delta \varphi }I_{|n|p}(rr^{\prime }/2s)=\frac{1}{p}\sum_{k=0}^{p-1}e^{\frac{rr^{\prime }}{2s}\cos (\frac{\Delta\varphi }{p}+\frac{2\pi k}{p})}\ .  \label{Kaa}
\end{equation}
The corresponding heat kernel reads,
\begin{equation}
{\mathcal{K}}(x,x^{\prime };s)=\frac{2e^{-sm^{2}}}{(4s\pi )^{D/2}}\sinh\left( \frac{zz^{\prime }}{2s}\right) \sum_{k=0}^{p-1}e^{-\frac{{\mathcal{V}}%
_{k}}{4s}}\ ,
\end{equation}
where
\begin{equation}
{\mathcal{V}}_{k}=-\Delta t^{2}+\Delta \mathbf{x}^{2}+z^{2}+z^{\prime}{}^{2}+r^{2}+r^{\prime }{}^{2}-2rr^{\prime }\cos \left( \Delta \varphi/p+2\pi k/p\right) \ .
\end{equation}

Finally, substituting the above function into (\ref{heat}), with the help of  Ref.~\refcite{Grad} we get,
\begin{equation}
G(x,x^{\prime })=\frac{m^{D-2}}{(2\pi )^{D/2}}\sum_{k=0}^{p-1}\left[f_{D/2-1}\left( m{\mathcal{V}}_{k(-)}\right) -f_{D/2-1}\left( m{\mathcal{V}}%
_{k(+)}\right) \right] \ ,  \label{g-special}
\end{equation}%
where
\begin{equation}
{\mathcal{V}}_{k(\mp )}=\left[ -\Delta t^{2}+\Delta \mathbf{x}^{2}+(z\mp z^{\prime })^{2}+r^{2}+r^{\prime }{}^{2}-2rr^{\prime }\cos \left( \Delta\varphi /p+2\pi k/p\right) \right] ^{1/2}\ .
\end{equation}
In (\ref{g-special}) and in what follows we use the notation
\begin{equation}
f_{\nu }(x)=K_{\nu }(x)/x^{\nu },  \label{fnu}
\end{equation}%
being $K_{\nu }(x)$ the Macdonald function. 

We can see that the Green function (\ref{g-special}) vanishes for $z$ or $z^{\prime }$ being equal to zero. Moreover, it can be presented as the sum of two
different contributions as shown below:
\begin{equation}
G(x,x^{\prime })=G_{\text{cs}}(x,x^{\prime })+G_{\text{b}}(x,x^{\prime })\ .
\label{GFdec}
\end{equation}
In (\ref{GFdec}), $G_{\text{cs}}(x,x^{\prime })$ coincides with the Green function for a massive scalar field in the absence of the boundary. It is divergent at the coincidence limit and the divergence comes from the $k=0$ term. As to $G_{\text{b}}(x,x^{\prime })$, it is a consequence of the boundary condition imposed on the field. This contribution is finite at the coincidence limit for points outside the boundary.

The formula for the Green function in the case of Neumann boundary condition is obtained from (\ref{g-special}) changing the sign of the terms with ${\mathcal{V}}_{k(+)}$. Consequently, the boundary induced parts in the Green function, $G_{\text{b}}(x,x^{\prime})$, for Dirichlet and Neumann scalars differ by the sign.

\subsection{General case}

For general case where $p=1/\alpha $ is not an integer number, the Green function can be expressed in an integral form. Substituting the heat kernel (\ref{heat2}) into (\ref{heat}), the Green function can be expressed in terms of a sum of boundary-free and boundary-induced parts, as exhibit below:
\begin{eqnarray}
G_{cs}(x,x') &=&\frac{p}{(4\pi )^{D/2}}\sum_{n=-\infty}^{+\infty }e^{in\Delta \varphi }\int_{0}^{\infty }dw\ w^{D/2-2}e^{-\frac{{\cal{V}_{(-)}}}{4}w-\frac{m^{2}}{w}}\ I_{|n|p}\left( rr'w/2\right) \nonumber\\
G_{b}(x,x') &=&-\frac{p}{(4\pi )^{D/2}}\sum_{n=-\infty}^{+\infty }e^{in\Delta \varphi }\int_{0}^{\infty }dw\ w^{D/2-2}e^{-\frac{\cal{V}_{(+)}}{4}w-\frac{m^{2}}{w}}\ I_{|n|p}\left( rr'w/2\right) \ \nonumber\\
\label{GFcsb}
\end{eqnarray}
with
\begin{equation}
{\cal{V}_{(\mp )}}=-\Delta t^{2}+\Delta \mathbf{x}^{2}+(z\mp z^{\prime})^{2}+r^{2}+r^{\prime }{}^{2}\ .  \label{Vpm}
\end{equation}
We can also verify that the Green function vanishes for $z$ or $z^{\prime}$ equal to zero.

\section{Vacuum Expectations Values}
This section will be devoted to the calculations of vacuum polarizations effects induced by the boundary. Two main calculations will be performed. The evaluation of the VEV of the field squared, in the first part, followed by the corresponding evaluation of the energy-momentum tensor.

\subsection{Calculation of $\langle\phi^2(x)\rangle$}

The evaluation of the VEV of the field squared is formally given by evaluating the Green function at the coincidence limit. However, in this analysis the complete Green function is given by the sum of the Green function is cosmic string spacetime in absence of boundary plus a boundary induced part. In this way we may write,
\begin{equation}
\langle \phi ^{2}\rangle =\langle \phi ^{2}\rangle _{\text{cs}}+\langle \phi^{2}\rangle _{\text{b}}\ .  \label{phi2sum}
\end{equation}%
Because $G_{\text{cs}}(x,x^{\prime })$ is divergent at the coincidence limit, the renormalization procedure is needed only for the first contribution of the above expression. The second contribution is finite at
the coincidence limit for points outside the hypersurface $z=0$. Because the VEV of the field squared in the cosmic string spacetime has been analyzed by many authors, here we are mainly interested in the analysis of the quantum effects induced by the boundary.

According to the previous section, we shall analyze the VEV of the field squared induced by the boundary for $p$ being an integer number in the first part, and for the general case in the second one.

\subsubsection{Special case}
Being $p$ an integer number, the VEV of the field squared induced by the boundary is given simply by taking the coincidence limit of $G_{\text{b}}(x^{\prime },x)$ given in (\ref{g-special}). The result is presented by the
expression
\begin{equation}
\langle\phi^{2}\rangle_{\text{b}}=-\frac{m^{D-2}}{(2\pi )^{D/2}}\sum_{k=0}^{p-1}f_{D/2-1}(2m\sqrt{z^{2}+r^{2}s_{k}^{2}})\ ,
\label{P2-special}
\end{equation}
where
\begin{equation}
s_{k}=\sin (\pi k/p)\ .  \label{sk}
\end{equation}

Below we present some asymptotic behaviors of (\ref{P2-special}) in different limiting regions:
\begin{itemize}
\item For $z\neq 0$ and for points far from the string, $r\gg |z|$, the dominant contribution comes from the $k=0$ component and to the leading order we get%
\begin{equation}
\langle \phi ^{2}\rangle _{\text{b}}\approx \langle \phi ^{2}\rangle _{\text{b}}^{(p=1)}=-\frac{m^{D-2}}{(2\pi )^{D/2}}f_{D/2-1}\left( 2m|z|\right) .
\label{phi20}
\end{equation}
\item In the limit $|z|\gg r$ and $m|z|\gg 1$, the leading order term provides an exponentially suppressed behavior below,
\begin{equation}
\langle \phi ^{2}\rangle _{\text{b}}\approx -\frac{p\ m^{(D-3)/2}e^{-2m|z|}}{2(4\pi )^{(D-1)/2}|z|^{(D-1)/2}}\ .  \label{phi2bsp}
\end{equation}
\item For a massless field, and in the limit $|z|\gg r$, we obtain
\begin{equation}
\langle \phi ^{2}\rangle _{\text{b}}\approx-\frac{p \ \Gamma (D/2-1)}{(4\pi )^{D/2}}\frac1{|z|^{D-2}} .  \label{phi2bsp0}
\end{equation}
\end{itemize}

\begin{figure}[tbph]
\begin{center}
\epsfig{figure=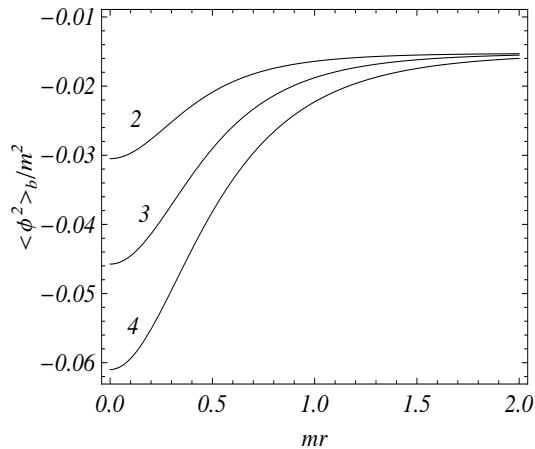, width=7.cm, height=6.cm}
\end{center}
\caption{This graph provides the behavior of $\langle \protect\phi^{2}\rangle _{\text{b}}/m^{2}$ in a 4-dimensional cosmic string spacetime as a function of $mr$ for $mz=0.5$ and for several values of the parameter $p$ (numbers near
the curves). }
\label{fig1}
\end{figure}

In Fig. \ref{fig1} we exhibit the behavior of (\ref{P2-special}) in the case of a $D=4$ cosmic string spacetime as a function of the dimensionless variable $mr $ for three distinct values of $p=2,\ 3,\ 4$ and for $mz=0.5$. We can see that the effect induced by the string becomes more relevant for larger values of $p$.

\subsubsection{General case}

In the case of $p$ being not an integer number, the VEV of the field squared induced by the boundary is obtained by taking the coincidence limit of $G_{\text{b}}(x^{\prime },x)$ given in (\ref{GFcsb}). After a change of the
integration variable it can be written as:
\begin{equation}
\langle \phi ^{2}\rangle _{\text{b}}=-\frac{pr^{2-D}}{(2\pi )^{D/2}}\int_{0}^{\infty }\ dy\ y^{D/2-2}e^{-\left( 2z^{2}/r^{2}+1\right)y-m^{2}r^{2}/(2y)}\ S_{p}(y)\ ,  \label{P2-general}
\end{equation}
with
\begin{equation}
S_{p}(y)=\sideset{}{'}{\sum}_{n=0}^{\infty }I_{np}(y)\ ,  \label{Sp}
\end{equation}
where the prime means that the term with $n=0$ should be halved.

In order to provide an explicit dependence of $\langle \phi ^{2}\rangle _{\text{b}}$ with $p$, we shall use the formula below:\footnote{The expression (\ref{SumForm}) is a special case of a more general formula derived in Ref.~\refcite{Beze10}.}
\begin{equation}
S_{p}(y)=\frac{1}{p}\sideset{}{'}{\sum}_{k=0}^{p_{0}}e^{y\cos (2\pi k/p)}-\frac{\sin (p\pi )}{2\pi }\int_{0}^{\infty }dx\frac{e^{-y\cosh x}}{\cosh(px)-\cos (p\pi )},  \label{SumForm}
\end{equation}
where $p_{0}$ is an integer number defined by $2p_{0}<p<2p_{0}+2$.

Substituting (\ref{SumForm}) into (\ref{P2-general}), the integration over $y $ can be explicitly performed and one finds
\begin{eqnarray}
\langle \phi ^{2}\rangle _{\text{b}} &=&-\frac{2m^{D-2}}{(2\pi )^{D/2}}\ \bigg[\sideset{}{'}{\sum}_{k=0}^{p_{0}}f_{D/2-1}(2m\sqrt{z^{2}+r^{2}s_{k}^{2}})  \notag \\
&&-\frac{p}{\pi }\sin (p\pi )\int_{0}^{\infty }dx\frac{f_{D/2-1}(2m\sqrt{z^{2}+r^{2}\cosh ^{2}x})}{\cosh (2px)-\cos (p\pi )}\bigg].  \label{phi2b}
\end{eqnarray}%
It can be seen that for integer values $p$ this result reduces to the expression (\ref{P2-special}).

\begin{figure}[tbph]
\begin{center}
\epsfig{figure=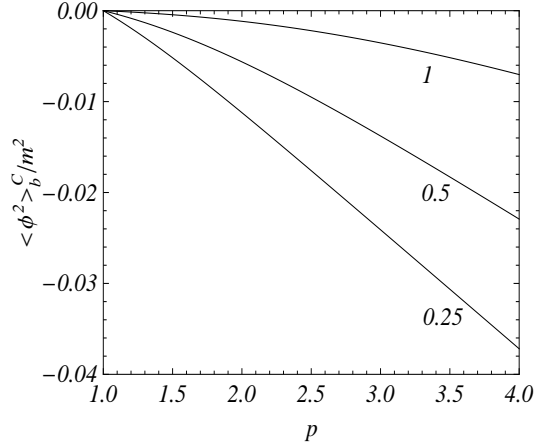, width=7cm, height=6.cm }
\end{center}
\caption{The topological part in the boundary induced VEV of the field squared, $\langle \protect\phi ^{2}\rangle _{\text{b}}^{\text{C}}/m^{2}$, in a 4-dimensional cosmic string spacetime as a function of $p$ for $mz=0.5$
and for several values of $mr$ (numbers near the curves). }
\label{fig2n}
\end{figure}

Defining
\begin{equation}
\langle \phi ^{2}\rangle_b=\langle \phi ^{2}\rangle_b^{(p=1)}+\langle \phi ^{2}\rangle_b^{\text{C}}\ ,  \label{phi2BC}
\end{equation}
in Fig. \ref{fig2n} we exhibit $\langle \phi ^{2}\rangle _{\text{b}}^{\text{C}}/m^{2}$ as a function of the parameter $p$ for $mz=0.5$ and for several values of $mr$ (numbers near the curves) in a 4-dimensional spacetime ($D=4$). Note that for the first term in the right-hand side of (\ref{phi2BC}), for $mz=0.5$ one has $\langle \phi ^{2}\rangle _{\text{b}}^{(p=1)}\approx-0.0152m^{2}$.

\subsection{Calculation of $\langle T_{\mu\nu}(x)\rangle$}
Following the same line of investigation, here we are interested to calculate the contribution induced by the boundary on the VEV of the energy-momentum tensor. Similar to the case of the field squared, the energy-momentum tensor is presented in the decomposed form,
\begin{equation}
\langle T_{\mu\nu}\rangle =\langle T_{\mu\nu}\rangle _{\text{cs}}+\langle T_{\mu\nu}\rangle _{\text{b}},  \label{EMTdec}
\end{equation}%
where $\langle T_{\mu\nu}\rangle _{\text{cs}}$ corresponds to the geometry of the cosmic string without boundaries. In order to evaluate the boundary induced part we shall use the following expression:
\begin{equation}
\langle T_{\mu\nu}\rangle _{\text{b}}=\lim_{x^{\prime }\rightarrow x}\partial_{\mu^{\prime }}\partial_{\nu}G_{\text{b}}(x,x^{\prime })+\left[ \left( \xi -{1}/{4}\right) g_{\mu\nu}\Box -\xi(\nabla_\mu\nabla_\nu+R_{\mu\nu})\right] \langle\phi ^{2}\rangle _{\text{b}}\ .  \label{mvevEMT}
\end{equation}%
For the spacetime under consideration, $R_{\mu\nu}=0$.

Here we shall investigate only the cases of $p$ be an integer number. Moreover, only two components will be separately examined: the energy-density, $\langle T_0^0\rangle _{\text{b}}$, and the pressure along $z-$axis, $\langle T_z^z\rangle _{\text{b}}$. Due to the presence of the boundary the boost invariance along the $z$-axis is lost and, as a consequence, $\langle T_z^z\rangle_B\neq\langle T_{0}^{0}\rangle_B$.

\subsubsection{The energy-density}
Let us start with the energy-density:
\begin{eqnarray}
\label{ED}
\langle T^0_0(x)\rangle_b&=&-\frac{m^{D/2}}{(4\pi)^{D/2}}\sum_{k=0}^{p-1}\frac1{u_k^{D/2+1}}\left\{[u_k^2+(4\xi-1)[(z^2+r^2s_k^4)(D-1)\right.\nonumber\\
&-&\left.s_k^2(2z^2+r^2(1+s_k^2))]]\frac{K_{D/2}(2mu_k)}{u_k}+2m(4\xi-1)(z^2+r^2s_k^4)\right.\nonumber\\
&&\left.\times K_{D/2-1}(2mu_k)\right\} \ ,
\end{eqnarray} 
where $u_k=\sqrt{z^2+r^2s_k^2}$ and $s_k=\sin(\pi k/p)$.

\begin{figure}[tbph]
\begin{center}
\begin{tabular}{cc}
\epsfig{figure=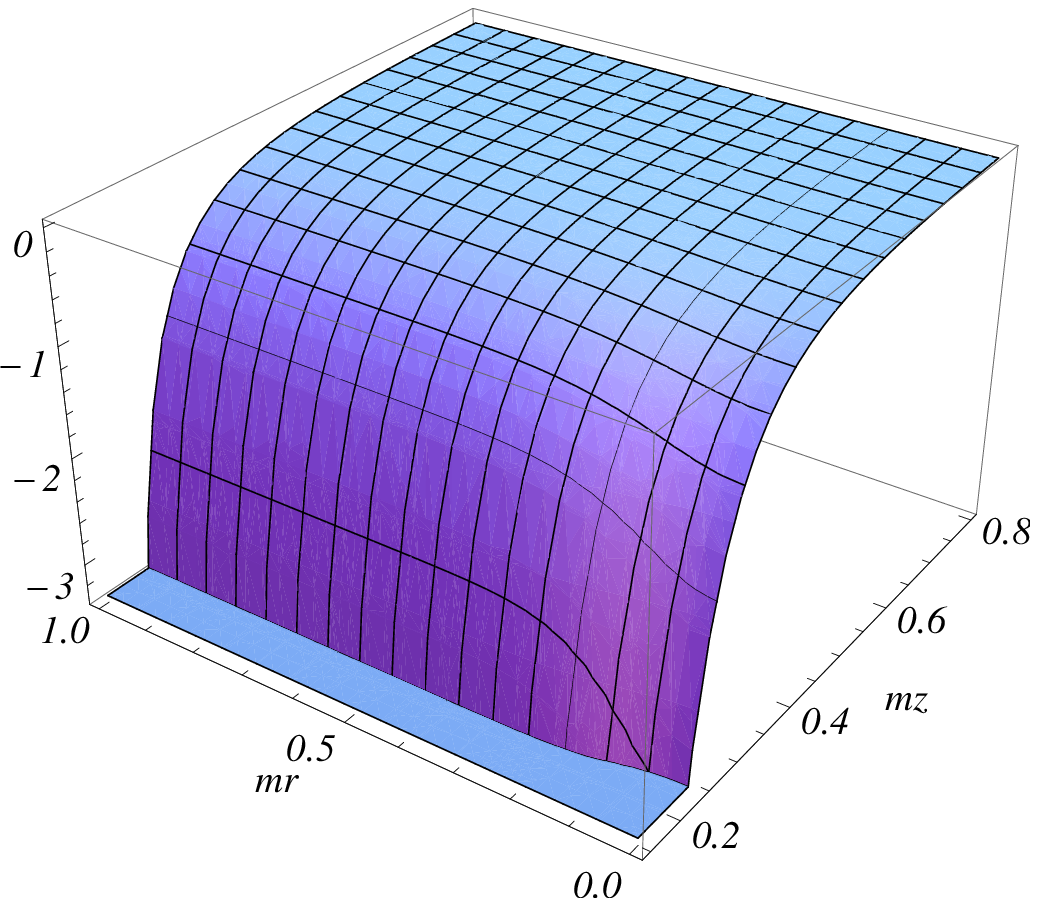, width=5.5cm, height=5.5cm} & \quad %
\epsfig{figure=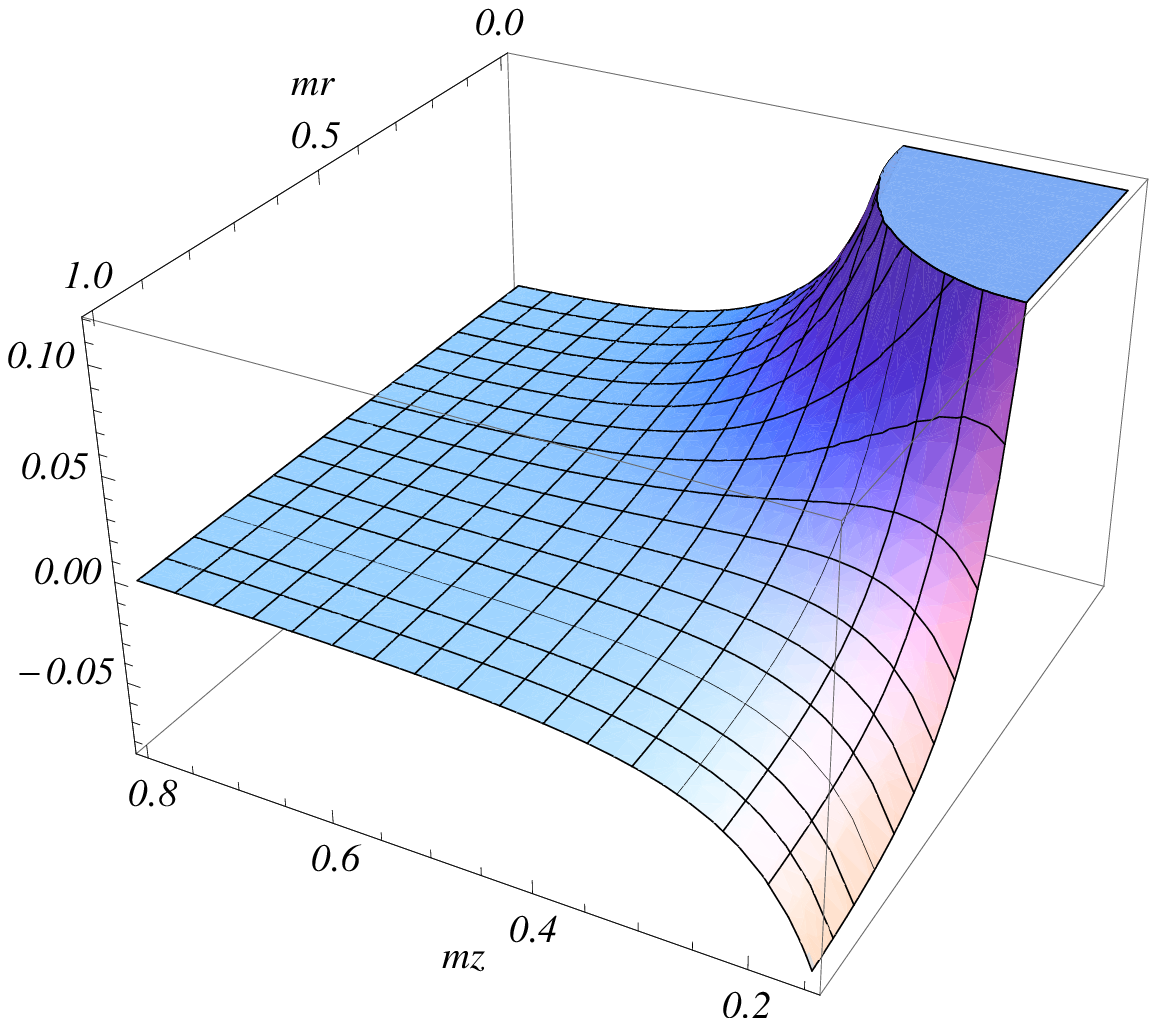, width=5.5cm, height=5.5cm}%
\end{tabular}%
\end{center}
\caption{The boundary induced part in the VEV of the energy density, $\langle T^{00}\rangle _{\text{b}}/m{^{4}}$, is exhibited for minimally coupled (left plot) and conformally coupled (right plot) scalar fields in a 4-dimensional spacetime as a function of dimensionless variables $mr$ and $mz$ for $p=3$.}
\label{fig3}
\end{figure}

In Fig. \ref{fig3} we exhibit the VEV of the energy density induced by the boundary for a minimally coupled, $\xi=0$, in the left plot, and conformally coupled, $\xi=1/6$, in the right plot, massive scalar fields in a 4-dimensional spacetime, as a function of dimensionless variables $mr$ and $mz$ for $p=3$. We can see that the energy density crucially depends on the curvature coupling parameter $\xi $.

\subsubsection{The pressure along the $z-$axis}

The pressure along the $z-$axis is important for the discussion about the pressure on the boundary presented in the following. The complete expression is:
\begin{eqnarray}
\label{Pressure}
\langle T^z_z(x)\rangle_b&=&\frac{m^{D/2}}{(4\pi)^{D/2}}(1-4\xi)\sum_{k=1}^{p-1}\frac{s_k^2}{u_k^{D/2+2}}\left\{[r^2s_k^2(D-2)-2z^2]K_{D/2}(2mu_k)\right.\nonumber\\
&+&\left.2u_kmr^2s_k^2K_{D/2-1}(2mu_k)\right\} \ ,
\end{eqnarray} 

The vacuum effective pressure on the boundary, given by $P=\langle T_{3}^{3}\rangle _{\text{b},z=0}$, is finite for points $r\neq 0$ and reads,
\begin{eqnarray}
\label{Pressure1}
P&=&\frac{m^{D/2}}{(4\pi r)^{D/2}}(1-4\xi)\sum_{k=1}^{p-1}s_k^{2-D/2}\left[(D-2)K_{D/2}(2mrs_k)\right.\nonumber\\
&+&\left.2mrs_kK_{D/2-1}(2mrs_k)\right] \ .
\end{eqnarray}
Note that the dependence of the pressure on the curvature coupling parameter appears in the form of the factor $(1-4\xi )$. For a massless field, the above expression reduces to
\begin{eqnarray}
\label{Pressure2}
P=(D-2)\frac{(1-4\xi )\Gamma (D/2)}{2(4\pi )^{D/2}r^{D}}\ \sum_{k=1}^{p-1}s_{k}^{2-D} \ .
\end{eqnarray}
In particular, for $D=4$ this formula is explicitly evaluated by using $\sum_{k=1}^{p-1}\sin^{-2}(\pi k/p)=\frac{p^2-1}{3}$. The result is:
\begin{equation}
\label{Pressure3}
P=\frac{1-4\xi }{48\pi ^{2}r^{4}}(p^{2}-1) \ . 
\end{equation}
For the both minimally and conformally coupled scalar fields the corresponding effective pressure is positive. Although (\ref{Pressure3}) has been obtained for integer value of $p$, it is an analytical function of this parameter, and by analytical continuation it remains valid for any value of $p$.

\begin{figure}[tbph]
\begin{center}
\begin{tabular}{cc}
\epsfig{figure=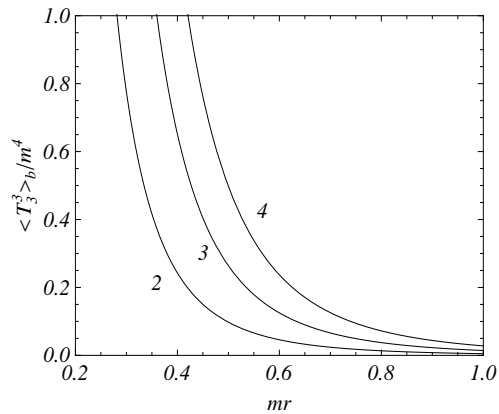, width=6.5cm, height=5.5cm}
\end{tabular}%
\end{center}
\caption{The normal vacuum stress on the boundary, $P$, for a minimally coupled scalar field in $D=4$ as a function of $mr$ for separate values of the parameter $p$ (numbers near the curves).}
\label{fig4}
\end{figure}

In Fig. \ref{fig4} we present the pressure on the boundary, $P$, for a minimally coupled scalar field in $D=4$ as a function of $mr$ for separate values of the parameter $p$ (numbers near the curves). 

\section{Conclusion}
 In this paper we have analyzed the effects induced by a flat boundary on the VEVs of the field squared and the energy-momentum tensor associated with a massive scalar field in a higher-dimensional cosmic string spacetime. The condition imposed on the field at the boundary is Dirichlet one. For a scalar field with Neumann boundary condition, the corresponding formulas for the VEVs of the field squared and the energy-momentum tensor are obtained from those given above by changing the sign of the boundary induced parts. In the presence of the boundary, the VEVs are given as a sum of two terms: the first one due to the cosmic string itself in the absence of boundary and the second terms induced by the boundary. Because the analysis of the VEVs associated with scalar fields in a pure higher-dimensional cosmic string spacetime have been developed in the literature, in this paper we were more interested to investigate the contribution induced by the boundary. Two distinct situations were considered: first one when the parameter which codifies the presence of the string is the inverse of an integer number, and in the second for general values of this parameter. For the first case, the Green function is expressed in terms of a finite sum of the Macdonald functions, while for the second one, only an integral representation can be provided.

For integer values of the parameter $p$, the boundary induced part in the VEV of the field squared is given by (\ref{P2-special}) and this part is always negative. For general case of the parameter $p$, we have provided the integral representation (\ref{phi2b}).

Another important local characteristic of the vacuum state is the VEV of the energy-momentum tensor. Similar to the field squared, the vacuum energy-momentum tensor is decomposed into pure topological and boundary induced parts. In the analysis of VEV of the energy-momentum tensor, only integer value of $p$ has been considered in this paper.\footnote{The complete analysis of VEV of the energy-momentum tensor has been developed in Ref. \refcite{Mello2} adopting a different approach.} Two components of the VEV of the energy-momentum tensor have been analyzed, the energy-density given by (\ref{ED}) and the pressure along the $z-$axis, given by (\ref{Pressure}). 

The normal vacuum force acting on the plate is determined by the component $\langle T_{z}^{z}\rangle _{\text{b}}$ evaluated on the boundary. This force is given by the expression (\ref{Pressure1}) for massive fields, and by (\ref{Pressure2}) for massless ones. Note that for a flat boundary in Minkowski spacetime the normal stress vanishes and the effective force in the geometry under consideration is induced by the presence of the string. For massless field in a $4-$dimensional cosmic string spacetime the pressure is given by (\ref{Pressure3}). 

The analytical results obtained in this paper explicitly show that the presence of a cosmic string increases the vacuum polarization effects induced by the boundary when compared with the corresponding ones in Minkowski spacetime. Regarding to the behavior of the VEV of the field squared, energy-density and the normal pressure on the boundary, some numerical results have been exhibited for a four dimensions, analyzing several different dependences. As our main conclusion we may say that the analytical results and the plots provide how relevant is the the presence of the cosmic string on the VEVs induced by the flat boundary.

\section*{Acknowledgments}
E.R.B.M. thanks Conselho Nacional de Desenvolvimento Cient\'{\i}fico e Tecnol\'{o}gico (CNPq) for partial financial support. A.A.S. thanks the organizers of the $8^{th}$ Alexander Friedmann International Seminar on Gravitation and Cosmology and CAPES for a support.



\begin{thebibliography}{99}
\bibitem{Kibble} T. W. Kibble, {\it J. Phys. A} {\bf 9}, 1387 (1976).
\bibitem{V-S} A. Vilenkin and E. P. S. Shellard, {\it Cosmic Strings and Other Topological Defects} (Cambridge University Press, Cambridge, England, 1994).
\bibitem{scalar} B. Linet, {\it Phys. Rev. D} {\bf 35}, 536 (1987).
\bibitem{scalar1} A. G. Smith, in {\it Symposium on the Formation and Evolution of Cosmic String}, edited by G. W. Gibbons, S. W. Hawking and T. Vachaspati (Cambridge University Press, Cambridge, England, 1989).
\bibitem{scalar2} P. C. Davies and V. Sahni, {\it Class. Quantum Grav.} {\bf 5}, 1 (1987).
\bibitem{scalar3} T. Souradeep and V. Sahni, {\it Phys. Rev. D} {\bf 46}, 1616 (1992).
\bibitem{scalar4} M. E. X. Guimar\~aes and B. Linet, {\it Class. Quantum Grav.} {\bf 10}, 1665 (1993).
\bibitem{ferm} V. P. Frolov and E. M. Serebriany, {\it Phys. Rev. D} {\bf 15}, 3779 (1287).
\bibitem{ferm1} B. Linet, {\it J. Math. Phys.} {\bf 36}, 3694 (1995). 
\bibitem{ferm2} E. S. Moreira Jnr., {\it Nucl. Phys. B} {\bf 451}, 365 (1995).
\bibitem{ferm3} V. B. Bezerra and N. R. Khusnutdinov, {\it Class. Quantum Grav.} {\bf 23}, 3449 (2006).
\bibitem{Cas} V. M. Mostepanenko and N. N. Trunov, {\it The Casimir Effect and Its Applications} (Clarendon, Oxford, 1997); M. Bordag, G. L. Klimchitskaya, U. Mohidden and V. M. Mostepaneko, {\it Advances in the Casimir effects} (Oxford University Press, Oxford, 2009)
\bibitem{Mello} E. R. Bezerra de Mello, V. B. Bezerra, A. A. Saharian and A. S. Tarloyan, {\it Phys. Rev. D} {\bf 74}, 025017 (2006).
\bibitem{Brev95} I. Brevik and T. Toverud, {\it Class. Quantum Grav.} {\bf 12}, 1229 (1995).
\bibitem{Mello1} E. R. Bezerra de Mello, V. B. Bezerra and A. A. Saharian, {\it Phys. Lett. B} {\bf 645}, 245 (2007).
\bibitem{Aram1} E. R. Bezerra de Mello, V. B. Bezerra, A. A. Saharian and A. S. Tarloyan, {\it Phys. Rev. D} {\bf 78}, 105007 (2008).
\bibitem{Beze10} E. R. Bezerra de Mello, V. B. Bezerra, A. A. Saharian and V. M. Bardeghyan, {\it Phys. Rev. D} {\bf 82}, 085033 (2010); S. Bellucci, E. R. Bezerra de Mello and A. A. Saharian, {\it Phys. Rev. D.} {\bf 83}, 085017 (2011).
\bibitem{Grad} I. S. Gradshteyn and I. M. Ryzhik, {\it Table of Integrals, Series and Products} (Academic Press, New York, 1980).
\bibitem{Pru} A. P. Prudnikov, Yu. A. Brychkov, and O. I. Marichev, {\it Integrals and Series} (Gordon and Breach, New York, 1986), Vol. 2.
\bibitem{Jean} J. Spinelly and E. R. Bezerra de Mello, {\it JHEP} {\bf 09}, 005 (2008).
\bibitem{Mello2} E. R. Bezerra de Mello and A. A. Saharian, {\it Class. Quantum Grav.} {\bf 28}, 145008 (2011) . 
 
\end{thebibliography}
\end{document}